%% file: PICOLON_NaI_Fin.tex
\documentclass{ptephy_v1}

\preprintnumber{XXXX-XXXX} 

\usepackage{amsmath}
\usepackage{graphicx}
\usepackage{wrapfig}
\usepackage{multirow}

\begin{document}

\title{Development of highly radiopure NaI(Tl) scintillator for PICOLON dark matter search project}


\author{K.~Fushimi$^{*}$}
\affil{Department of Physics, Tokushima University, 2-1 Minami Josanajima-cho, Tokushima city, Tokushima , 770-8506, Japan
\email{kfushimi@tokushima-u.ac.jp}}
\author{Y.~Kanemitsu}
\affil{Graduate School of Integrated Arts and Sciences, Tokushima University,
1-1 Minami Josanajima-cho, Tokushima city, Tokushima , 770-8502, Japan}
\author[2]{S.~Hirata \footnote{Present address: Otsuka Pharmaceutical Factory, Inc. Tokushima Itano Factory, Matsunami, Itano-cho, 
Itano-gun, Tokushima, 779-0195, Japan}}
\author{D.~Chernyak}
\affil{Department of Physics and Astronomy, University of Alabama, Tuscaloosa, Alabama 35487, USA 
and Institute for Nuclear Research of NASU, 03028 Kyiv, Ukraine}
\author{R.~Hazama}
\affil{Department of Environmental Science and Technology, Osaka Sangyo University, 3-1-1 Nakagaito, 
Daito city, Osaka, 574-8530, Japan}
\author{H.~Ikeda}
\affil{Research Center for Neutrino Science, Tohoku University, 6-3 Aramaki Aza Aoba, Aobaku, Sendai city, 
Miyagi, 980-8578, Japan }
\author{K.~Imagawa}
\affil{I.~S.~C. Lab.~, 5-15-24 Torikai Honmachi, Settsu city, Osaka, 566-0052, Japan}
\author{H.~Ishiura}
\affil{Department of Physics, Graduate School of Science, Kobe University, 1-1 Rokkodai-cho, Nada-ku, Kobe city, Hyogo, 657-8501, Japan}
\author{H.~Ito}
\affil{Institute for Cosmic Ray Research, The University of Tokyo, 5-1-5 Kashiwanoha, Kashiwa city, 
Chiba, 277-8583, Japan}
\author{T.~Kisimoto}
\affil{Department of Physics, Osaka University, 1-1 Machikaneyama-cho,
Toyonaka city,  Osaka 560-0043, Japan}
\author{A.~Kozlov}
\affil{National Research Nuclear University ``MEPhI'' (Moscow Engineering Physics Institute), Moscow, 115409, Russia}
\author[8,11]{Y.~Takemoto}\affil{Kavli Institute for the Physics and Mathematics of the Universe (WPI),
5-1-5 Kashiwanoha, Kashiwa city, Chiba, 277-8583, Japan}
\author[6]{K.~Yasuda}
\author{H.~Ejiri}
\affil{Research Center for Nuclear Physics, Osaka University, 10-1 Mihogaoka Ibaraki city, Osaka, 567-0042, Japan}
\author[5]{K.~Hata}
\author{T.~Iida}
\affil{Faculty of Pure and Applied Sciences, University of Tsukuba, 1-1-1 Tennoudai, Tsukuba city, 
Ibaraki, 305-8577, Japan}
\author[5,11]{K.~Inoue}
\author[5,11]{M.~Koga}
\author[5,11]{K.~Nakamura\footnote{Present address: Butsuryo College of Osaka, 3-33 Ohtori Kitamachi, Nishi ward, Sakai city, Osaka, 
593-8328, Japan }}
\author[1]{R.~Orito}
\author[12]{T.~Shima}
\author[12]{S.~Umehara}
\author[9]{S.~Yoshida}


\begin{abstract}%
The highly radiopure NaI(Tl) was developed to search for particle candidates of dark matter.
The optimized methods were combined to reduce various radioactive impurities.
The $^{40}$K was effectively reduced by the re-crystallization method.
The progenies of the decay chains of uranium and thorium were reduced by appropriate resins.
The concentration of natural potassium in NaI(Tl) crystal was reduced down to 20~ppb.
Concentrations of alpha-ray emitters were successfully reduced by appropriate selection of resin.
The present concentration of thorium series and $^{226}$Ra were
$1.2\pm1.4$~$\mu$Bq/kg and $13\pm4$~$\mu$Bq/kg, respectively.
No significant excess in the concentration of $^{210}$Pb was obtained, and the upper limit was 5.7~$\mu$Bq/kg at 90\% C.~L.
The achieved level of radiopurity of NaI(Tl) crystals makes construction of a dark matter detector possible.
\end{abstract}

\subjectindex{H20, F40}

\maketitle

\section{Introduction}
\input{intro_Fin.tex}

\section{Purification methods}
\input{purification_Fin.tex}

\section{Data taking and analysis}
\input{daq_Fin.tex}
\input{conclusion_Fin.tex}

\section*{Acknowledgement}
We acknowledge the support of the Kamioka Mining and Smelting Company. This work was
supported by JSPS KAKENHI Grant No. 26104008, 19H00688, 20H05246, and Discretionary expense of the president of Tokushima University.
This work was also supported by the World Premier International Research Center Initiative (WPI Initiative).
We acknowledge Profs.~H.~Sekiya and A.~Takeda of ICRR University of Tokyo, and Prof.~Y.~Takeuchi of Kobe University 
for continuous encouragement and fruitful discussions.


\bibliographystyle{ptephy}
\bibliography{sample}
%

%
%
%
%
%
%

\end{document}

%% file: intro_Fin.tex
Producing a high-sensitivity radiation detector for dark matter is currently the most crucial subject.
Many groups are trying to find a WIMPs signal by various methods and target nuclei.
Although a large volume liquid xenon (LXe) detector set stringent limits on
the existence of WIMPs dark matter candidates\cite{Aprile2019, Abe2019, Akerib2019},
DAMA/LIBRA group reported the significant signal of WIMPs
by applying highly radiopure
and large volume NaI(Tl) scintillator.
They observed the annual modulating signal of WIMPs rate due to the earth's revolution around the Sun \cite{Freese1988}.
The event rate and deposited energy of WIMPs modulate with their maximum at the beginning
of June and a minimum at the beginning of December.
The DAMA/LIBRA group developed a highly radiopure NaI(Tl) scintillator whose total mass was
250 kg.
They reported a significant annual modulating amplitude between 2 keV$_{\mathrm{ee}}$ and
6 keV$_{\mathrm{ee}}$, where keV$_{\mathrm{ee}}$ is the unit of energy scale calibrated by
the kinetic energy of electron \cite{Bernabei2013}.

Many groups are trying to verify the DAMA/LIBRA's result by  NaI(Tl) scintillator \cite{Fushimi2020, Amar2020, Adhikari2019, Antonello2020}.
The COSINE group started the low background measurement from 2018 by high-purity NaI(Tl) scintillator with its total 106 kg \cite{Adhikari2019}.
The event rate in the low energy region was about $2\sim3$ times larger than the one of DAMA/LIBRA.\@
The higher background was due to the intrinsic radioactive contamination, for example, $^{210}$Pb and $^{40}$K.\@
The beta rays from $^{210}$Pb, $^{210}$Bi, and $^{40}$K make a continuous background in the energy region of interest (below 10 keV). 
The Compton continuum due to 1462 keV gamma-ray emitted after electron capture of $^{40}$K also contributes to the background. 
These background events obscure the dark matter signals. 
Recently, the COSINE reported the improved NaI(Tl) crystal, which was produced by re-crystallization \cite{Park2020}.
They proved the effectiveness of the re-crystallization method to reduce the radioactive contamination in NaI(Tl).

We have tried further purification by applying a cation exchange resin,
in addition to the re-crystallization method.
The appropriate selection of the resin enabled the effective reduction of the lead ion in the NaI solution.
We developed an extremely high-purity NaI(Tl) crystal whose concentration of $^{210}$Pb is less than 6 $\mu$Bq/kg.

%% file: purification_Fin.tex
\subsection{Re-crystallization method}
The re-crystallization (RC) method helps remove the radioactive impurities that are well soluble in water.
The solubility of NaI in water is 75.14 at 100 $^{\circ}$C, and
it decreases 64.76 at 25 $^{\circ}$C.\@
The solubility is defined as the mass of solute in
100 g of water solution.
The difference in the solubility between 100 $^{\circ}$C and 25 $^{\circ}$C enables us to get the pure NaI sediment.

The COSINE group showed the significant $^{40}$K suppression in the RC method\cite{Shin2018}.
Potassium in the aqueous solution of sodium iodide forms a well soluble
potassium iodide (KI) or potassium hydroxide (KOH).
The solubility of KI and KOH in water are as high as
59.7 and 54.2 at 25 $^{\circ}$C, respectively, as shown in Table \ref{tb:solu}.
Since the concentration of potassium is much lower than its solubility, potassium ions remain in the water 
during the NaI RC process.
The COSINE group reported the effectiveness of the RC method \cite{Shin2018, Park2020}.
\begin{table}[!h]
\centering
\caption{The solubility of NaI, KI, KOH, and PbI$_{2}$ in water \cite{KagakuBinran}. }
\label{tb:solu}
\begin{tabular}{l|rrrr} \hline
Temp.($^{\circ}$C) & 0 & 20 & 25 & 100 \\ \hline
NaI & 61.54 & 64.1 & 64.76 & 75.14 \\
KI & 56.0 & 59.0 & 59.7 & 67.4 \\
KOH & 49.2 & 53.8 & 54.2 & 64.0 \\
PbI$_{2}$ & 0.041 & 0.060 & 0.076 & 0.43 \\ \hline
\end{tabular}
\end{table}


We verified effectiveness of the RC method by measuring the $^{40}$K activity in NaI 
before and after this procedure. 
The activity of $^{40}$K in the NaI was determined using an ultra-low background HPGe detector 
which is installed at the KamLAND underground facility of the Kamioka neutrino observatory.
The HPGe detector was constructed by joint efforts of Kavli IPMU (Tokyo Univ.) and 
RCNS (Tohoku Univ.) research groups. 
 More information about the HPGe detector and underground clean room facility can be found in 
 \cite{Kozlov2020, Kozlov2019}.
We found an apparent reduction of gamma-ray intensity by the RC method.
The initial radioactivity of $^{40}$K in the original (non-purified) NaI powder was in the range of 
$18\sim32$ mBq/kg at 90\% CL.\@
After a single RC cycle, the activity of $^{40}$K was reduced to $0.2\sim10$ mBq/kg at 90\% CL.\@

We prepared 100 $^{\circ}$C NaI saturated solution in a bottle filled with pure nitrogen.
The bottle was slowly cooled down to room temperature to avoid adhering of 
NaI crystals to the bottle's surface. 
The NaI crystals were separated from the water solution by suction filtration. 
The filtration was done in a glove box flushed with pure nitrogen to prevent contact with air containing radon.

\subsection{Resin method}
The COSINE group reported the re-crystallization worked well to reduce $^{210}$Pb \cite{Shin2018, Park2020}.
Our results also showed a significant reduction by double and triple RC methods, as listed in Table~\ref{tb:Limits}.
Nevertheless, we need a higher reduction to get a pure NaI(Tl) crystal below a few tens of $\mu$Bq/kg.
We tried to combine another strategy to remove Ra and Pb ions.

We extensively searched for the source of $^{210}$Pb and found it was not only 
$^{222}$Rn in pure water but also $^{210}$Pb in original NaI powder.
We combined several methods to remove both $^{222}$Rn and $^{210}$Pb from the water 
solution of NaI.\@
The gaseous $^{222}$Rn in water was removed by bubbling the pure water with pure nitrogen gas.
The longer bubbling, typically one hour, gives a significant reduction.
We tried to catch Pb in the water solution of NaI by both cation exchange resin and crown ether.

We searched for the best combination of the resins by processing the NaI aqueous solution, which was initially added 4.9 ppm of lead ion.
We tested four resins delivered by two companies. 
The resin A is suitable to remove strontium, uranium, and lead ions. 
The resin B is designed to remove the lead ion. 
The resins C and D, which another company delivers, are applied to remove uranium and iron ions.

We measured the concentration of the Pb ion in the sample by ICP-MS,
Agilent 7900 at Osaka University and Agilent 7700 at Osaka Sangyo University.
The measured values of the reduction factor of lead ion are listed in Table \ref{tb:samples}.
According to the result, we selected the resin B to reduce lead ion.
In addition to resin B, we apply the resins A to reduce not only
$^{210}$Pb but also U, Ra, and Th.
\begin{table}[!h]
\centering
\caption{The reduction factors for concentration of Pb ions in the NaI water solution 
achieved by use of various resins.}
\label{tb:samples}
\begin{tabular}{l|lrr} \hline
ID & Sample & Reduction factor \\ \hline
1 & Resin A & 1/34 \\
2 & Resin B & 1/64 \\
3 & Resin C & 1/14 \\
4 & Resin D & 1/3 \\ \hline
\end{tabular}\end{table}

\subsection{Crystallization}
We took necessary precautions to avoid contacts between the room's air and 
NaI powder during drying process.
We dried the purified NaI solution by a rotary evaporator.
We selected a flask made of synthetic silica glass to avoid the possible pollution from it.
We filled pure nitrogen gas into the flask to break the vacuum when the drying finished.
Additional drying was done by using a vacuum oven before growing NaI(Tl) crystal.

The large volume NaI(Tl) crystal was produced by the Bridgeman-Stockbarger technique.
The purified and dried NaI powder and thallium iodide were filled into a high-purity graphite crucible.
Certain details describing purification of the crucible material can be found in \cite{Kozlov2020, Kozlov2019}.
The crucible was installed in a quartz vessel filled with inert gas at the center of the furnace.
The crucible was heated to 700 $^{\circ}$C to melt the NaI powder, and it was slowly moved to crystallize.
The crystallization takes at least two weeks of cooling and annealing to make a transparent,
colorless, and large single crystal without any inclusion.
The melted NaI does not interact and infiltrate the pure graphite crucible because of the special coating on the surface
of the crucible.
In the end of crystal formation the NaI(Tl) ingot detaches from the crucible. 
We see no evidence of interaction between the crucible and melted NaI.\@ 
We think that there is no significant pollution of the NaI(Tl) ingot occurs by the crucible material.

%% file: daq_Fin.tex
\subsection{Detector construction}
The original NaI(Tl) ingot was used to make a NaI(Tl) crystal shaped as a cylinder 
(a 76.2~mm in diameter and 76.2~mm long). 
An optical window made of synthetic quartz with 10~mm in thickness was attached on one side of the crystal edge.
Other surfaces of the crystal were covered with an enhanced specular sheet (ESR) made of polyethylene.
The whole crystal was encapsulated into a 4~mm-thick black acrylic container.  
We used the acrylic housing to minimize absorption of low energy gamma and X-rays 
emitted by external calibration sources.

An acrylic housing enables a low energy calibration down to 6.4~keV$_{\mathrm{ee}}$.
The K$_{\beta}$ X-rays emitted after the electron capture of $^{133}$Ba are between 34.9~keV and 35.8~keV with total intensity of 11.5 \% \cite{TOI}.
After the photoelectronic effect on iodine, the ionized iodine emits its K$_{\alpha}$ X-ray, 28.6~keV and 28.3 keV \cite{TOI}.
 If these K$_{\alpha}$ X-rays escape from the crystal a residual energy of approximately 
 6.4 keV is deposited in a NaI(Tl) scintillator.
This low energy peak is useful to test the stability of low energy calibration.

Acrylic housing has a risk of moisture permeability.
We tested deliquesce when the NaI(Tl) detector was installed in the shield with pure nitrogen.
The NaI(Tl) detector has not suffered any deliquescence for at least 1.8~years.

We made several trials of purification. The details of the NaI(Tl) detectors are listed in Table~\ref{tb:detector}.
Hereafter, we identify the NaI(Tl) detector by crystal ID shown in Table~\ref{tb:detector}.
\begin{table}[!ht]
\centering
\caption{The list of NaI(Tl) detectors and purification techniques used in their production.}
\label{tb:detector}
\begin{tabular}{l|l} \hline
Crystal ID & Purification method \\ \hline
\#24 & Resin D+ Radium reduction resin \\
\#68 & ResinD + cation exchange resin\\
\#71 & RC twice \\
\#73 & RC triple \\
\#85 & RC twice \& resins A+B \\ \hline
\end{tabular}
\end{table}

\subsection{Data acquisition system}
We measured activity of radionuclides remained in the crystals \#68, \#71 and \#73 using 
an ultra low-background setup \cite{Kozlov2019} at the KamLAND underground facility.
The radiopurity of two other crystals (\#24 and \#85) was measured in a surface laboratory 
at Tokushima University.

In Kamioka, the DAQ system consists of VME based MoGURA electronics \cite{Terashima2008} 
produced by Tokyo Electron Device Ltd.\@
Several supplementary NIM modules were
used to reject majority of noise pulses from a photomultiplier tube (PMT) \cite{Kozlov2020}. 
The MoGURA on-board flash analog-to-digital converters (FADC) can be used to record up to 
10 $\mu$sec long waveforms.
Signal selection as well as rejection of the remaining noise pulses from PMT were done by 
offline pulse shape analysis (PSD).

In the surface laboratory, the current pulse from a PMT was divided into three routes and digitized by CAMAC ADC \cite{Ichihara2003}.
The one was used for a trigger. 
We introduced one current signal of the PMT to a charge-sensitive ADC (CSADC: REPIC RPC-022) channel through a 200~ns cable delay.
The CSADC integrates the total charge of the current pulse for 1 $\mu$s.
The other current signal was introduced into the other ADC channel directly to integrate it partially.

The PSD was performed to extract the alpha-ray event.
The decay time of the NaI(Tl) for an alpha-ray, $\tau_{\alpha}$ is 190~ns,
on the other hand, the one for a beta or gamma-ray, $\tau_{\beta}$ is
230~ns\cite{Bernabei2008}.
This difference enables clear discrimination of an alpha-ray and others.
The parameter $R$ was calculated as
\begin{equation}
R\equiv \frac{\int^{t_{2}}_{t_{1}}I(t)dt}{\int^{t_{2}}_{t_{0}}I(t)dt},
\end{equation}
where $I(t)$ was the input current.\@
The upper limit and the lower limits of the integration are defined as $t_{2}$, $t_{0}$, and $t_{1}$, respectively.
The present values of $t_{0}$, $t_{1}$, and $t_{2}$ were listed in Table \ref{tb:integ}.
\begin{table}[ht]
\centering
\caption{Integration parameters $t_{0,1,2}$ nsec in two measurement
systems.}
\label{tb:integ}
\begin{tabular}{lrrr} \hline
& $t_{0}$ & $t_{1}$ & $t_{2}$ \\ \hline
MoGURA & 0 & 200 & 1200 \\ 
CAMAC & 0 & 200 & 1000 \\ \hline
\end{tabular}
\end{table}

\subsection{Measurement of $^{40}$K in underground laboratory}
We analyzed dependence of the concentration of $^{40}$K in the NaI(Tl) detectors 
\#68, \#71, and \#73 on various purification methods listed in Table \ref{tb:detector}.
\begin{figure}[!h]
\centering
\includegraphics[width=0.45\textwidth]{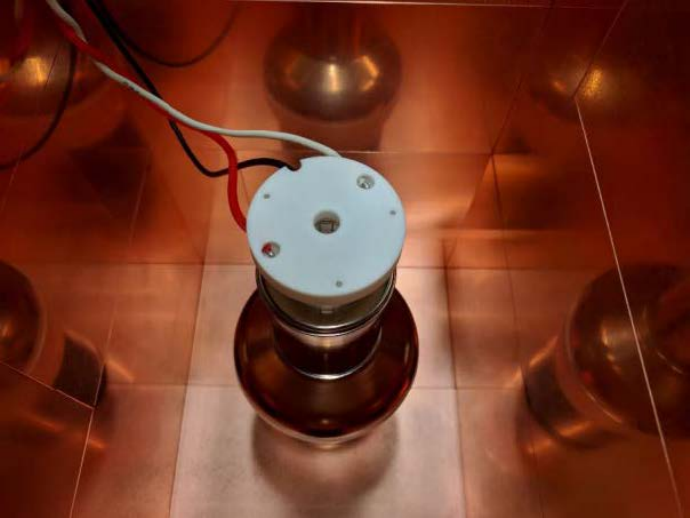}
\includegraphics[width=0.44\textwidth]{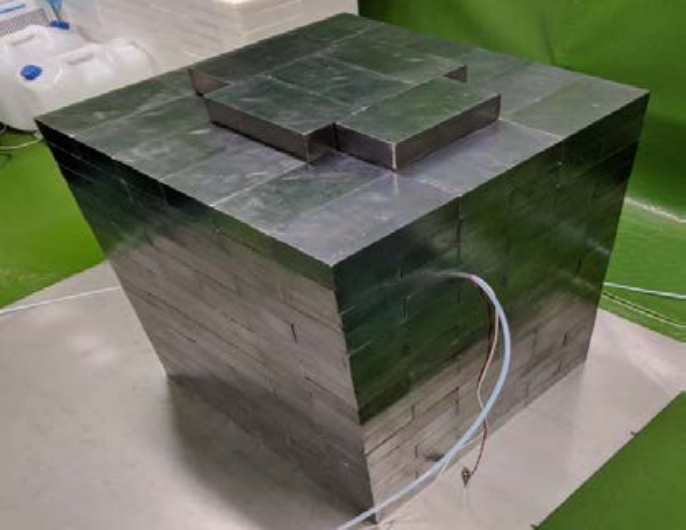}
\caption{The detector and the shield of the present measurement in Kamioka.}
\label{fg:I73}
\end{figure}

We measured the concentration of $^{40}$K in the NaI(Tl) crystal at Kamioka Underground laboratory, Tohoku University.
The $^{40}$K decays both beta decay (89.26\%) and electron capture (10.72\%).
We need a low background environment to measure the tiny amount of radioactivity because of the enormous external background.
The Kamioka Underground laboratory is suitable to reduce the background from cosmic rays.
To suppress environmental gamma-rays, the detectors were placed inside of an ultra-low background 
passive shielding made of a $15\sim20$~cm-thick lead and a 5~cm-thick 99.99\% pure copper layers.
Structure and composition of the passive shielding were also reported in the earlier publications \cite{Kozlov2020,Kozlov2019}.
Figure \ref{fg:I73} shows crystal \#73 surrounded by the copper shielding layer.
An ultra-low background PMT, Hamamatsu Photonics 4-inch R13444X \cite{Kozlov2020},
was attached to the optical window of the NaI(Tl) crystal by an optical grease.
A pure nitrogen gas was supplied into the shield to purge out air containing radon.

\begin{figure}[ht]
\centering
\includegraphics[width=0.45\linewidth]{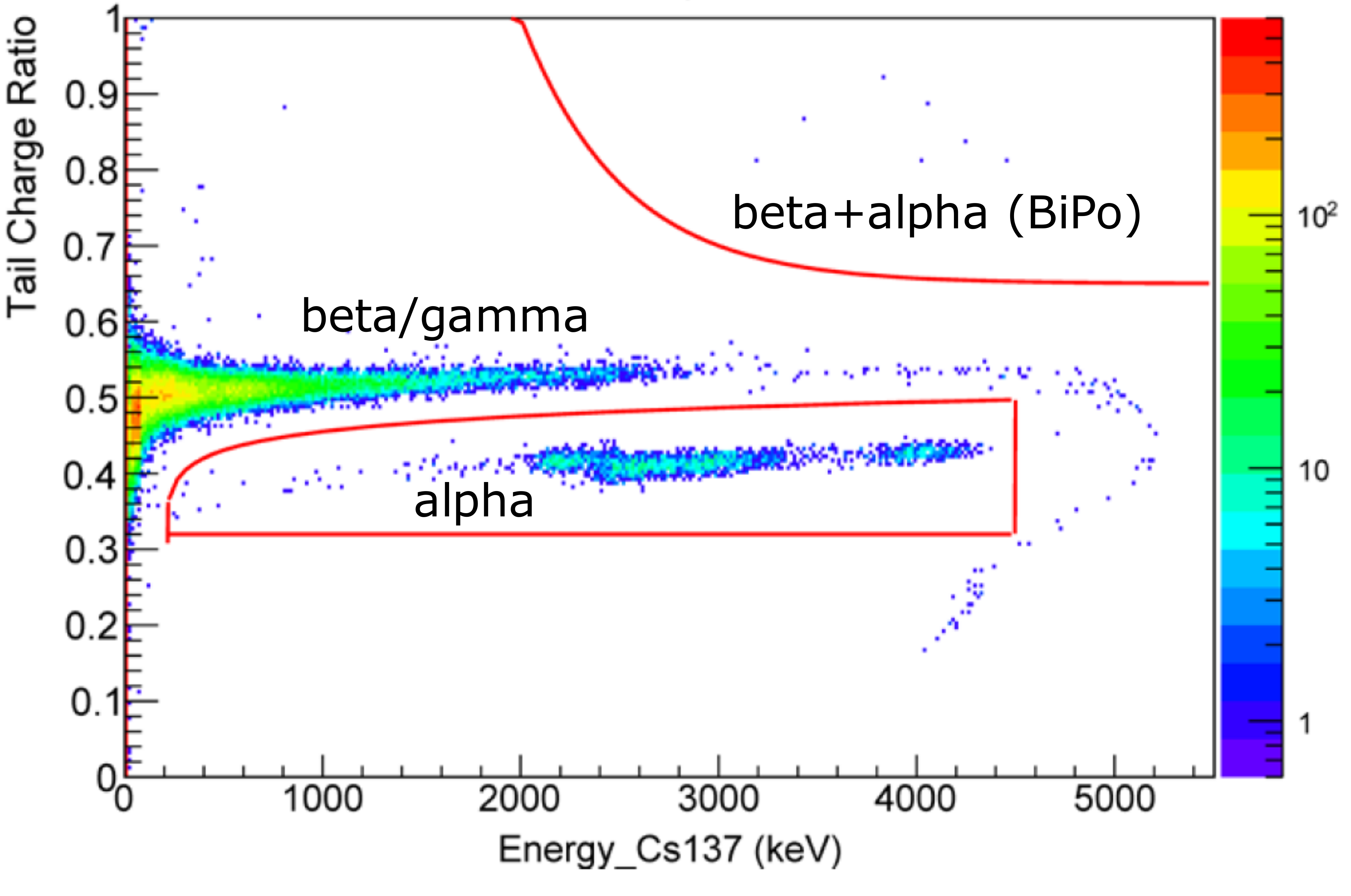}
\includegraphics[width=0.45\linewidth]{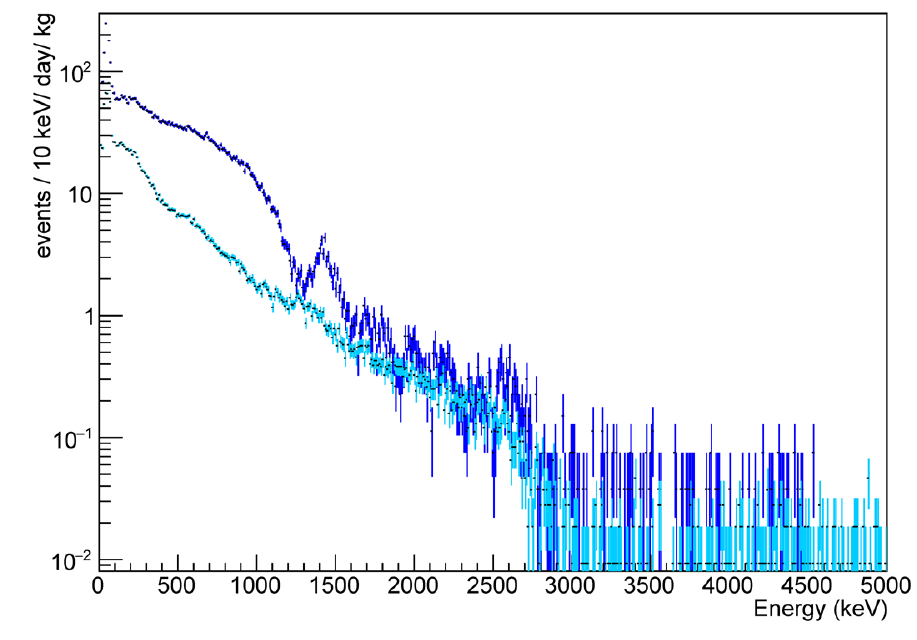}
\caption{Left: PSD analysis result in case of crystal \#71.
Right: The energy spectra of low-background measurement.
Blue: Result by \#68. Cyan: Result by \#71.}
\label{fg:I6871}
\end{figure}
The MoGURA data acquisition system recorded the pulse shape of the PMT output.
The pulse shape analysis was done, as shown in the left panel of Figure \ref{fg:I6871}.
The Energy-Ratio distribution makes three loci.
The PSD distribution shows the significant distortion above 
4000~keV$_{\mathrm{ee}}$ due to the saturation of PMT output.
The beta/gamma-ray energy spectrum was obtained without distortion to check the existence of the beta-ray ($E_{\beta\mathrm{max}}=1311.09$ keV) and gamma-ray ($E_{\gamma}=1461$ keV) \cite{TOI}.
We selected the beta- and the gamma-ray events by PSD analysis.
The energy spectra observed by the crystals \#68 and \#71
are shown in the right panel of Figure \ref{fg:I6871}.
There was a clear difference in the energy spectra between crystals \#68 and \#71.
There was a prominent continuum below 1.3 MeV$_{\mathrm{ee}}$ of crystal \#68 due to the beta-rays of $^{40}$K.\@
The prominent beta-ray spectrum insisted that the $^{40}$K was contained in the NaI(Tl) crystal of \#68.
The concentration of $^{\mathrm{nat}}$K in the crystal \#68 was calculated as 130 ppb by Geant 4.10 Monte Carlo simulation,
assuming the natural abundance of $^{40}$K in natural potassium is
$0.0117$\%.
On the other hand, there were no significant components of both
beta-ray and gamma-ray from $^{40}$K in the energy spectrum of the crystal \#71.
The beta-ray event rate in crystal \#71 was at least 1/6 smaller than in crystal \#68.
It corresponds to the concentration of $^{\mathrm{nat}}$K was less than 20~ppb at 90\% C.~L.

The reproducibility and the effectiveness of the triple RC method were tested by ingot \#73.
The upper limit of $^{\mathrm{nat}}$K was 30 ppb at 90\% C.~L.
We found that the double RC method was enough to reduce potassium contamination.
We confirmed the double RC method's effectiveness by later ingots \#76 and \#83.

\begin{figure}[ht]
\centering
\includegraphics[width=0.45\linewidth]{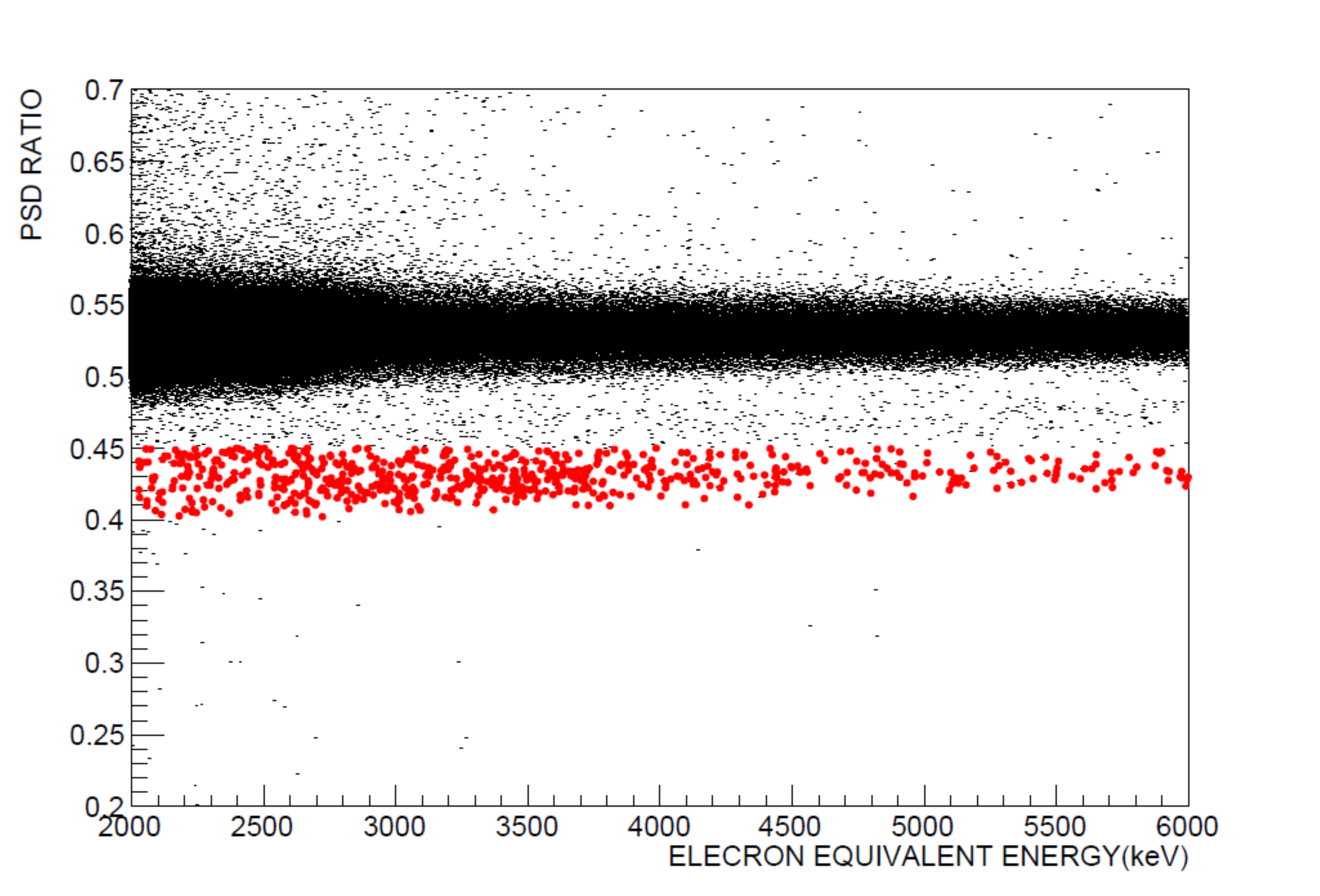}
\hspace{0.05\linewidth}
\includegraphics[width=0.45\linewidth]{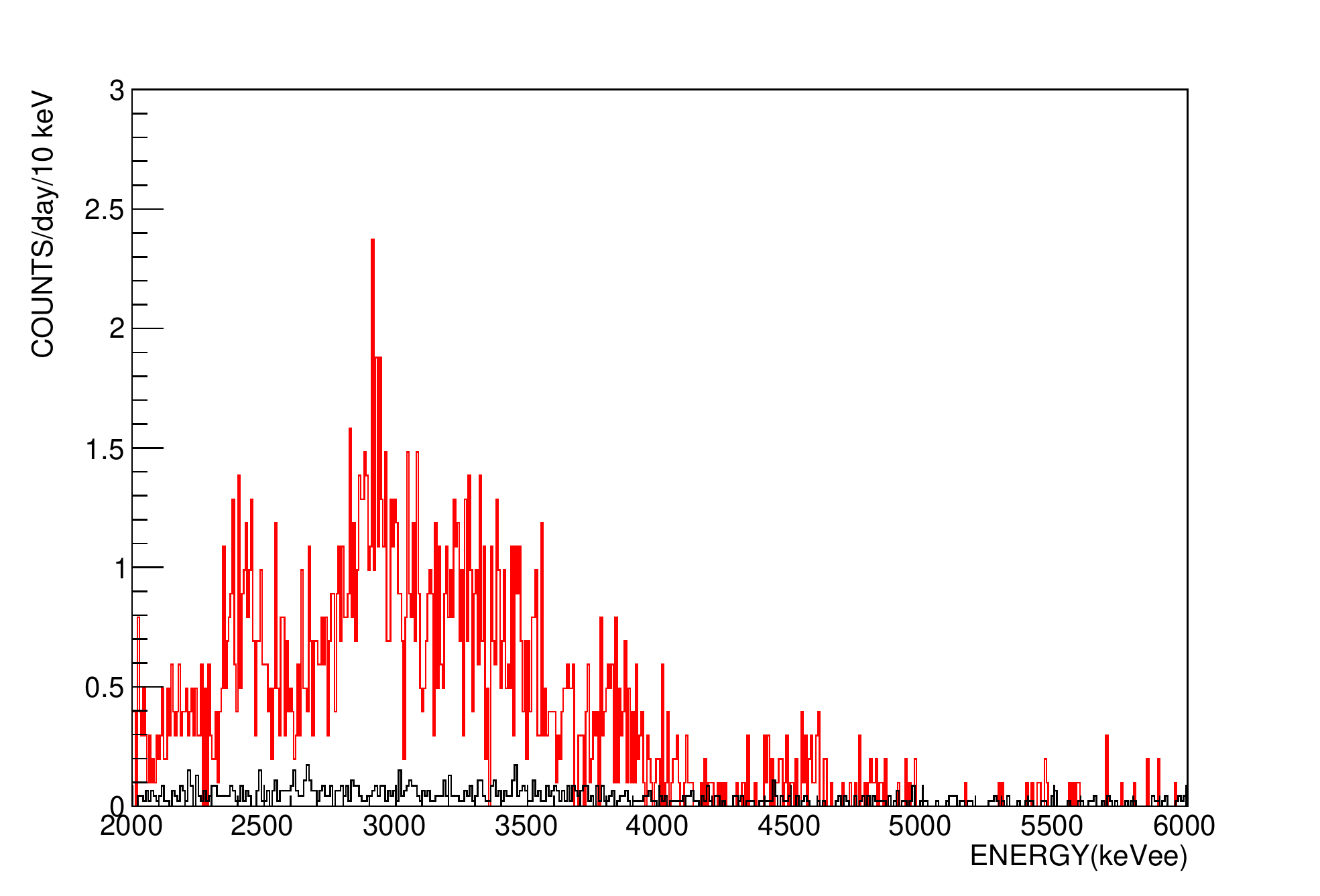}
\caption{Left: The PSD plot with crystal \#85.
The red points indicate the selected alpha-ray events.
Right: Comparison of alpha-ray energy spectrum between crystal \#24 (red) and crystal \#85 (black).}
\label{fg:PSD_I85}
\end{figure}
\subsection{Measurement of alpha-rays in surface laboratory}
We measured the alpha concentration in purified NaI(Tl) crystals.
After establishing the potassium reduction, we optimized the reduction method of alpha-ray emitters.
The present crystal \#85 was made by the combined method of double RC and resins.
We also measured the alpha-rays in \#24 to verify the reduction.

We accumulated the data for the live-time of
36.33~day$\times 1.279$~kg for \#85 and 11.2~day$\times 1.279$~kg for \#24.
The result of the PSD analysis with the crystal \#85 is shown in Figure \ref{fg:PSD_I85}.
It shows the faint concentration in the region $0.4<R<0.45$.
A large locus around $R=0.53$ is the events of beta-rays, gamma-rays, and cosmic-rays.
We compared the energy spectrum of alpha-rays with one of our previous best NaI(Tl) crystal, \#24 \cite{fushimi2014kamlandpico}.
The right panel of Figure \ref{fg:PSD_I85} is the energy spectra taken by crystals \#24 (red) and \#85 (black).
The horizontal axis is the electron equivalent energy.
There was a considerable improvement in the event rate in the present work.
The crystal \#24 contains $163\pm33$ $\mu$Bq/kg of $^{226}$Ra,
$149\pm11$ $\mu$Bq/kg of $^{228}$Th, and $86\pm25$ $\mu$Bq/kg of $^{210}$Pb.
We analyzed the constituent of the alpha-ray spectrum of crystal \#85 as described below.
\begin{wrapfigure}{r}{0.5\linewidth}
\centering
\includegraphics[width=\linewidth]{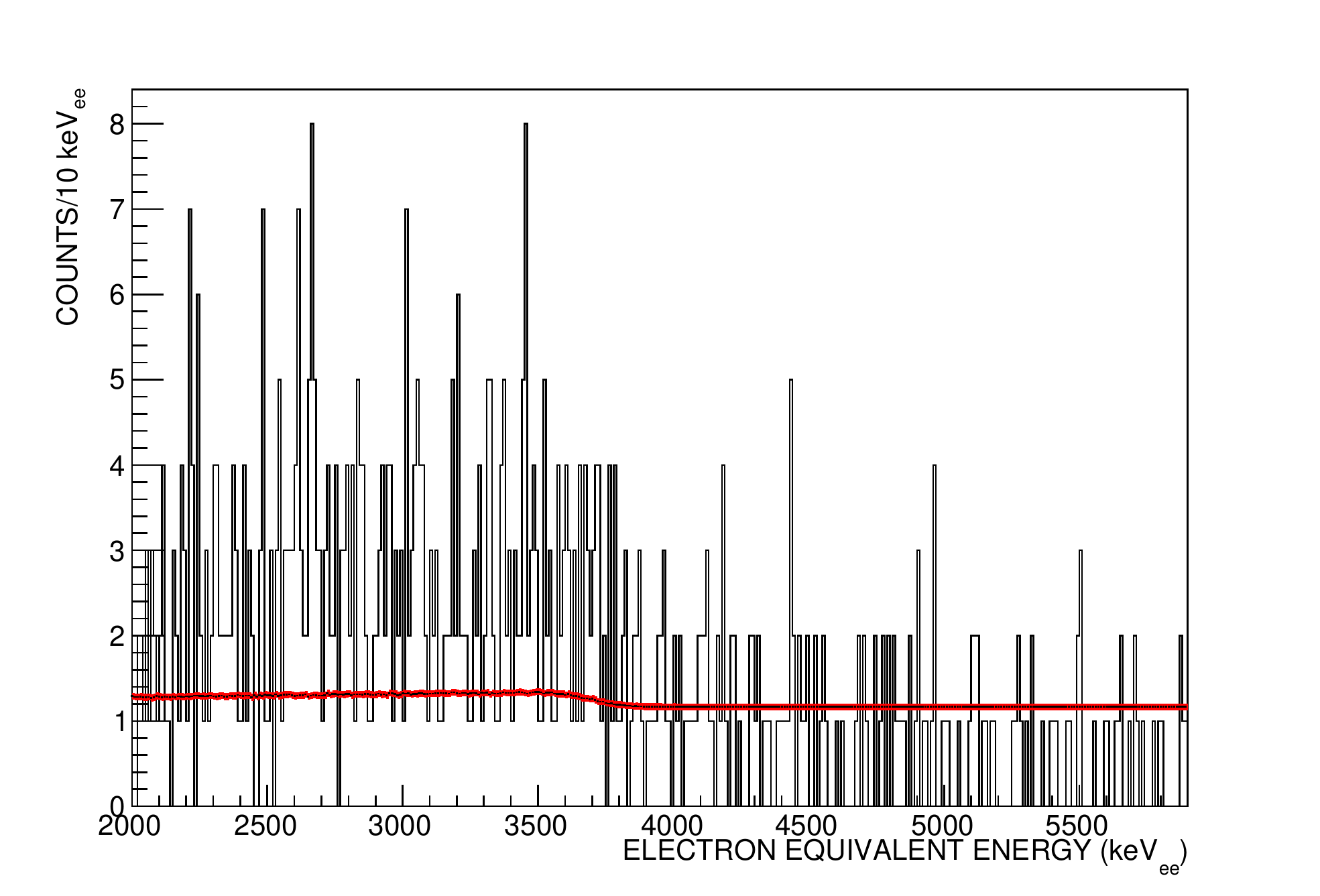}
\caption{The energy spectrum of alpha-ray taken by
crystal \#85.}
\label{fg:I85Alpha}
\end{wrapfigure}

The alpha-ray energy spectrum consists of two components (see Figure \ref{fg:I85Alpha}).
One is the lower energy region (below 4000 keV$_{\mathrm{ee}}$), and the other is the higher energy region (above 4000 keV$_{\mathrm{ee}}$).
Both components have no prominent structure.
We considered that the events above 4000 keV$_{\mathrm{ee}}$ are an accidental spread of $R$ of high energy cosmic rays.
We assumed there is a constant background of the cosmic-ray events below 4000 keV$_{\mathrm{ee}}$.
The energy spectrum below 4000 keV$_{\mathrm{ee}}$ consists of
constant cosmic-ray, alpha-ray emitted in the NaI(Tl) crystal, and
alpha-ray emitted on the ESR reflector sheet.

\begin{figure}[!ht]
\centering
\includegraphics[width=\linewidth]{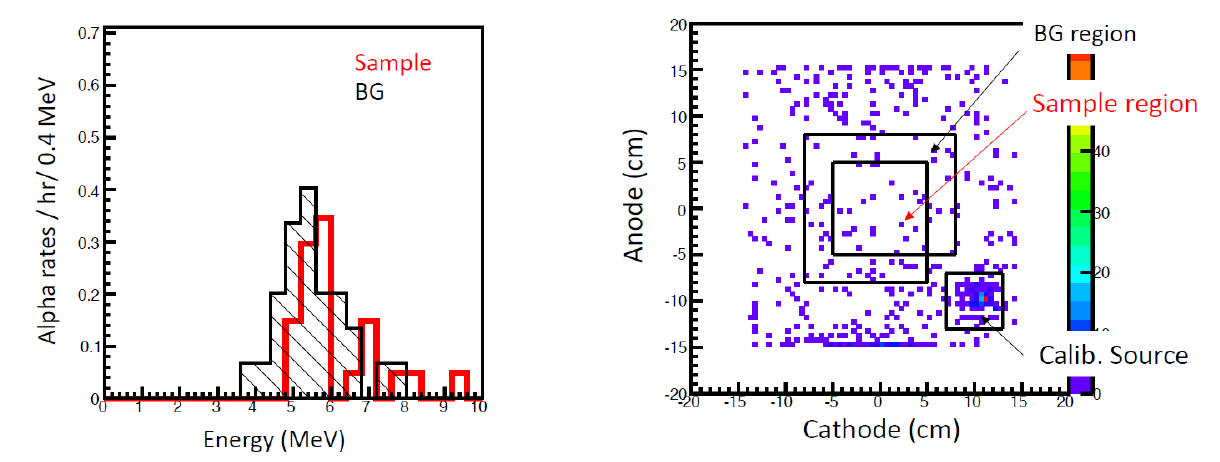}
\caption{Left: The energy spectra inside (red) and outside (black shaded) the sample region. Right: The distribution of emitting points of alpha-ray events. The ESR sheet is placed in the sample region. }
\label{fg:muPIC}
\end{figure}

We measured the surface concentration of alpha-ray emitters on the ESR sheet to determine the alpha-ray intensity.
We applied the low-background and position-sensitive alpha-ray tracker, $\mu-$PIC\cite{Hashimoto2020, Ito2020}.
No significant excess beyond the background was observed, as shown in Figure \ref{fg:muPIC}.
The upper limit of the surface contamination was calculated as $1.77 \times 10^{-3}$ alpha/hr/cm$^{2}$ (90\% C.~L.).

We simulated the energy spectrum from the ESR sheet by Geant4.10 Monte Carlo simulation \cite{Allison2016}.
We assumed the origin of the alpha-ray was $^{210}$Po because the ESR
sheet attracts the progeny of $^{222}$Rn.
$^{210}$Pb terminates the decay chain from $^{222}$Rn, and
accumulated $^{210}$Pb generates the $^{210}$Po.
A red line in Figure \ref{fg:I85Alpha} shows the calculated energy spectrum of $^{210}$Po from the ESR sheet.
There is significant excess in the event rate in the low energy region above the red line.
We concluded the excess is due to the alpha-rays emitted in the NaI(Tl) crystal.

The energy spectrum of alpha-ray from the NaI(Tl) crystal showed no prominent structure.
The event number of alpha ray was too small to perform further analysis such as time-correlation analysis. 
We calculated the yield by integrating the number of events instead of the peak fitting. 
We set the integration interval according to the quenching factor of alpha-ray and energy resolution.
The alpha-rays which have near energies make a cluster in the energy spectrum because of low energy resolution.
We analyzed the alpha-ray intensity according to the clusterized energy ranges was calculated according to the quenching factor of alpha-ray in NaI(Tl) scintillator \cite{fushimi2014kamlandpico}.
The energy listed in Table \ref{tb:ana} is the FWHM (Full Width Half-Maximum) interval of the clusterized peak.
The chemical symbol in the parenthesis indicates the decay chains of uranium and thorium.
The electron equivalent energy of alpha-ray, $E_{\mathrm{ee}}$, is
derived by
\begin{equation}
E_{\mathrm{ee}}=f_{\alpha}E_{\alpha}.
\end{equation}
Where $f_{\alpha}$ is the quenching factor of alpha-ray.
We determined $f_{\alpha}=0.58$ from the shape of the energy spectrum.

The alpha-ray of $^{224}$Ra lies between (C) and (D) clusters.
The alpha-rays of $^{224}$Ra contributes both event rates of (C) and (D) because of low energy resolution.
We divided two clusters at the peak energy of the alpha-ray of $^{224}$Ra.
The event rates of the clusters (C) and (D) contain half of the $^{224}$Ra events.
We confirmed the result of this analysis by the data from crystal  \#24.
The present analysis agrees with the result by peak fitting.

\begin{table}[!ht]
\centering
\caption{The RIs of alpha-ray emitters. The energy range (keV$_{\mathrm{ee}}$) and the event number in the regions are listed.
Detailed treartment of $^{224}$Ra in described in the text.}
\label{tb:ana}
\begin{tabular}{clrr} \hline
ID &RIs & Energy range& Events \\ \hline
(A) & $^{238}$U (U), $^{232}$Th (Th) & $2180\sim2550$ & $45\pm10$ \\
(B) & $^{234}$U (U), $^{230}$Th (Th), $^{226}$Ra (U) & $2580\sim2900$ & $60\pm11$ \\
(C) & $^{228}$Th (Th), $^{224}$Ra* (Th), $^{222}$Rn (U), $^{210}$Po (U) & $2970\sim3300$ & $44\pm10$ \\
(D) & $^{218}$Po (U), $^{212}$Bi (Th), $^{224}$Ra* (Th), $^{220}$Rn (Th) & $3300\sim3740$ & $66\pm12$ \\
(E) & $^{216}$Po (Th) & $3820\sim4043$ & $5\pm6$ \\ \hline
\end{tabular}
\end{table}

The concentration of radioactivity was calculated from the event numbers in Table \ref{tb:ana} and the total exposure 36.33 day$\times$1.279 kg.
We assumed that all the radioactivities of isotopes in the decay chain of thorium were in secular equilibrium.
The isotopes of uranium-series, $^{226}$Ra, $^{222}$Rn, $^{218}$Po, and $^{214}$Po were in secular equilibrium.
We did not include $^{214}$Po in the present analysis since we cannot take all the events of alpha-rays from $^{214}$Po.
The half-life of $^{214}$Po is as short as 164 $\mu$sec, and the dead-time of the present data acquisition system was as long as several hundreds of microsecond.
The concentration of $^{218}$Po was easily derived from the result of (D) as $14\pm4$ $\mu$Bq/kg.
The one of $^{226}$Ra was derived from the result of (B) as $13\pm4$ $\mu$Bq/kg; both values are consistent with each other.

No significant excess was found in the concentration of $^{210}$Po.
We set the upper limit on the concentration as 5.7 $\mu$Bq/kg at 90\%C.~L.

%% file: conclusion_Fin.tex
\section{Prospects}
\begin{wrapfigure}{r}{0.5\linewidth}
\vspace{-1cm}
\centering
\includegraphics[width=\linewidth]{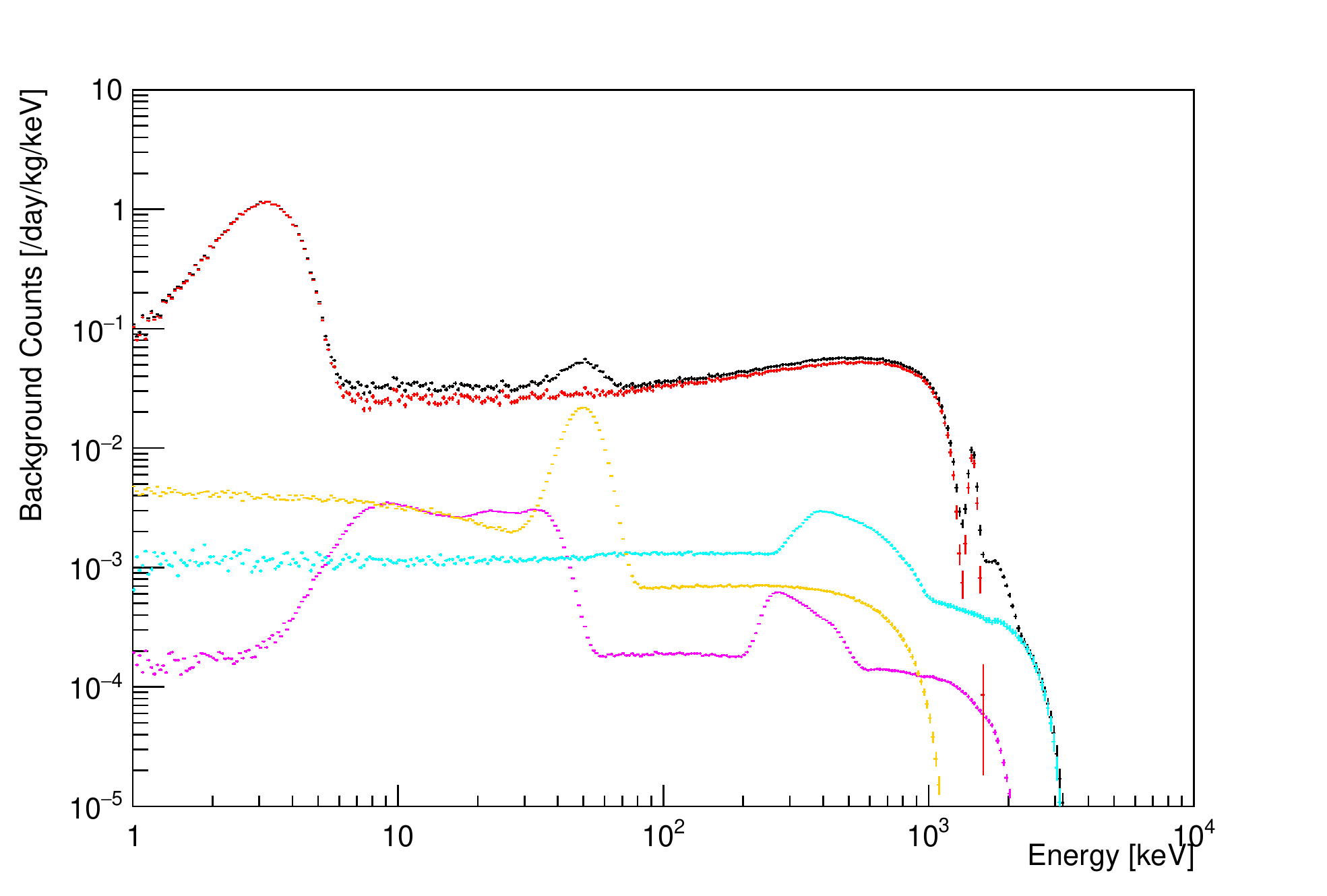}
\caption{The expected contribution of RIs from the NaI(Tl) crystal. Black: Total. Red: $^{40}$K.
Orange: $^{210}$Pb and $^{210}$Bi. Blue: $^{226}$Ra. Magenta: Th series RIs.}
\label{fg:sim}
\end{wrapfigure}
We successfully developed the highly-radiopure NaI(Tl) scintillator.
The severe background origins, $^{40}$K, $^{210}$Pb, and $^{226}$Ra, were successfully removed by combining the RC method and resin method.
All values of radioactive imurities in the crystal \#85 were below our goals.
\begin{table}[!t]
\centering
\caption{The concentration of $^{\mathrm{nat}}$K and alpha-ray emitters in NaI(Tl) scintillators
\cite{Kozlov2020}.
Units of $^{\mathrm{nat}}$K is in ppb, and others are in $\mu$Bq/kg.}
\label{tb:Limits}
\begin{tabular} {l|rrrr} \hline
ID of NaI/group & $^{\mathrm{nat}}$K & $^{226}$Ra & $^{210}$Pb & $^{232}$Th \\ \hline
\#68 & 120 & $57\pm7$ & 7500 & $8.4\pm2.4$ \\
\#71 & $<20$ & $120\pm10$ & 1500 & $6.8\pm0.8$ \\
\#73 & $<30$ & $44\pm7$ & 1300 & $7.2\pm0.8$ \\
\#85 & -- & $13\pm4$ & $<5.7$ & $1.2\pm1.4$ \\
Our goal & $<20$ & $<100$ & $<10$ & $<10$ \\ \hline
COSINE\cite{Park2020}
& $<42$ & $8\sim60$ & $10\sim420$ & $7\sim35$ \\
DAMA\cite{Bernabei2008}
& $<20$ & $8.7\sim124$ & $10\sim30$ & $2\sim31$ \\
\hline
\end{tabular}
\end{table}

We expect the highly sensitive verification of the DAMA/LIBRA experiment.
A large volume scintillator whose dimension is $12.7$ cm in diameter and $12.7$ cm in length is under construction.
The expected energy spectrum due to the intrinsic RIs are shown in Figure \ref{fg:sim}.
The next origin of the background is mainly the gamma-ray from external RIs. 
They will be selected and removed by anti-coincidence of arrayed NaI(Tl) detector system.

We will start the dark matter search experiment by large NaI(Tl) scintillators, PICOLON phase-I, from 2021 spring.
It consists of at least four modules of NaI(Tl) scintillator, whose total mass is 23.4 kg.
We plan the phase-II and phase-III with the total mass of 100 kg and 250 kg of NaI(Tl) crystal to verify and search for new fundamental processes in nuclear and particle physics.